\documentclass[runningheads]{llncs}
\usepackage[T1]{fontenc}
\usepackage{graphicx}
\usepackage{algorithmic}
\usepackage{textcomp}
\usepackage{caption}
\usepackage{tcolorbox}
\usepackage{xcolor,colortbl}
\usepackage{wrapfig}

\usepackage{makecell}
\usepackage{pifont}
\usepackage{makecell}
\usepackage{textcomp}
\usepackage{hyperref}
\usepackage[utf8]{inputenc}
\usepackage[T1]{fontenc}
\usepackage{multirow}
\usepackage{multicol}

\definecolor{lightgray}{rgb}{0.83, 0.83, 0.83}
\definecolor{lightorange}{rgb}{1, 0.8, 0.6}

\author{Rebeka Tóth$^{1}$, Tamas Bisztray$^{1}$, and László Erd\H{o}di$^{1}$
\thanks{*This work was not supported by any organization}
\thanks{$^{1}$R. Tóth, T. Bisztray and L. Erd\H{o}di is with the Faculty of Department of Informatics, University of Oslo, Norway.
        {\tt\small emails: rebekat at uio.no, tamasbi at uio.no, laszloe at uio.no}}%
}

\begin{document}
\title{LLMs in Web Development: Evaluating LLM-Generated PHP Code Unveiling Vulnerabilities and Limitations}
%
%
\author{Rebeka Tóth\inst{1}\orcidID{0009-0000-9574-1896} \and
 Tamas Bisztray\inst{1}\orcidID{0000-0003-2626-3434} \and
László Erd\H{o}di\inst{1}\orcidID{0000-0002-4910-4228}}

\authorrunning{R. Tóth et al.}
%
\institute{University of Oslo, Norway$^1$\\
\email{rebekat@ifi.uio.no}, 
\email{tamasbi@ifi.uio.no},
\email{laszloe@ifi.uio.no}\\}
\maketitle              
\begin{abstract}
This study evaluates the security of web application code generated by Large Language Models, analyzing 2,500 GPT-4 generated PHP websites. These were deployed in Docker containers and tested for vulnerabilities using a hybrid approach of Burp Suite active scanning, static analysis, and manual review. Our investigation focuses on identifying Insecure File Upload, SQL Injection, Stored XSS, and Reflected XSS in GPT-4 generated PHP code. This analysis highlights potential security risks and the implications of deploying such code in real-world scenarios. Overall, our analysis found 2,440 vulnerable parameters. According to Burp's Scan, 11.56\% of the sites can be straight out compromised. Adding static scan results, 26\% had at least one vulnerability that can be exploited through web interaction. Certain coding scenarios, like file upload functionality, are insecure 78\% of the time, underscoring significant risks to software safety and security. To support further research, we have made the source codes and a detailed vulnerability record for each sample publicly available. This study emphasizes the crucial need for thorough testing and evaluation if generative AI technologies are used in software development.

\end{abstract}
\section{INTRODUCTION}

In software development, ensuring the correctness, safety, and security of the developed programs is crucial. Functionally correct code provides the anticipated result for every input it receives. The goal of safety is to create code that is fail-safe, guarding against accidental or unforeseen inputs that could lead to logically correct yet unwanted outputs. Software security refers to the software's robustness against external threats and intentional attacks \cite{ref_fetzer}. 
LLMs are increasingly utilized by developers for tasks such as code completion, generation, and translation \cite{ref_hou,ref_ross}. Despite their benefits, generative AI introduces novel risks due to its non-deterministic and unexplainable nature. These risks are coupled with the possibility of generating hallucinated outputs that can result in misinformation, unforeseen behaviors, inherent biases, and vulnerabilities in code generation \cite{ref_tihanyi}, thereby compromising software safety and security.
While most research on LLM-based code generation examines only task correctness \cite{ref_hou}, our study focuses on software safety and security aspects. In software development, generative AI is employed in two main ways: code completion and code generation. Code completion aims at context-aware, real-time coding suggestions without explicit user prompts. In contrast, code generation relies on explicit prompts to produce code blocks, aligning more with direct task automation and specific functionality requests. Our focus will be the latter case, giving neutral prompts without explicit request for secure code.

The PHP server-side scripting language remains integral to web development, powering over 75\% of websites \cite{ref_kyriakakis}. Its continual releases and updates ensure its ongoing relevance in the web development landscape \cite{ref_smart,ref_scam2022}.
Our objective is to assess the capability of the \texttt{gpt-4-0125-preview} model to generate secure, functional PHP code without requiring further modifications or supervision. We aim to address the following research questions:
\begin{itemize}
\item \textbf{RQ1}: What is the extent of vulnerabilities in the PHP code generated by gpt-4-0125-preview in a zero-shot setting?
\item \textbf{RQ2}: Does the generated PHP code achieve a level of complexity and realism suitable for real-life deployment?
\end{itemize}
Merely showing that LLMs can generate code containing vulnerabilities through a few real-life use-cases would not be sufficient to answer our research questions, and has been previously performed across different programming languages \cite{ref_tihanyi,ref_nehorai,ref_perry}.
Our paper presents the following original contributions:
\begin{itemize}
\item We developed ChatPHP-DB, an AI-generated PHP website dataset that consists of 2,500 standalone PHP programs with accompanying init.sql files. 
\item Classification of each program based on vulnerabilities identified through active scans, static scans, or manual verification, focusing on: Insecure File Upload, SQL injection, Stored XSS, and Reflected XSS.

\end{itemize}
We made the dataset and the vulnerability labelling publicly available on GitHub: \nolinkurl{https://github.com/Beckyntosh/ChatPHP}.
The remaining sections are structured as follows: Section~\ref{sec:motivation} discusses the inspiration for our work. Section~\ref{sec:related} overviews the related literature. Section~\ref{sec:methodology} outlines the approach we employed to create the dataset, and perform vulnerability detection. Section~\ref{sec:experiments} provides the results and an in-depth evaluation of our findings. Section~\ref{sec:limitations} overviews limitations related to our work. Finally, Section~\ref{sec:conclusions} concludes the paper with an outlook on possible future research directions.

\section{MOTIVATION} \label{sec:motivation}


We must consider whether explicitly requesting secure code when testing AI-based code generation would be a good strategy. In code completion scenarios, specifically requesting secure code is not feasible to begin with. Moreover, Tihanyi et al. demonstrated that even when vulnerabilities are specifically identified, LLMs commonly fail to rectify them adequately \cite{ref_tihanyi}. Given that tools like Copilot X use specialized GPT models for features such as autocompletion and code analysis, assessing the secure coding capabilities of the underlying model itself is crucial.
In a recent blog post, security researcher Nathan Nehorai reviewed several LLM-based code completion tools, looking at tasks like fetching files, comparing secret tokens, and handling password resets \cite{ref_nehorai}.
He pointed out that security risks often depend on the situation: what is secure in one setting could become a vulnerability when handling external user inputs. This kind of context is hard for AI to grasp, as it mainly learns from data patterns, making it difficult for the AI to distinguish between safe and risky practices.

\begin{figure}
    \centering
    \includegraphics[width=1\linewidth]{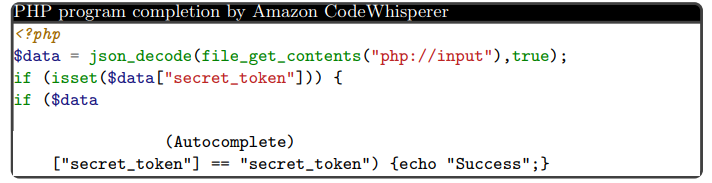}
    \caption{Vulnerable code completion example }
    \label{fig:phpvul}
\end{figure}

The second example in the article is highly instructive.
As shown in Listing \ref{fig:phpvul}, the suggested loose comparison for $secret\_token$ creates a vulnerability that an adversary can exploit, bypassing the check ``$==$''. When a JSON object is submitted where the value of $secret\_token$ is set as $True$ (a boolean value), the result of this loose comparison will also be $True$, a behaviour that persists even in newer versions of PHP.

Nehorai notes that even if specifically asking for a secure version, the tool failed to generate a code snippet incorporating a strict comparison ``$===$'', necessary against type juggling attacks.
Previous research has shown that LLMs are unreliable in fixing vulnerabilities even when explicitly requested \cite{ref_tihanyi}.
Therefore, our objective is to assess the performance of \texttt{gpt-4-0125-preview} across different programming scenarios and contexts, specifically focusing on its propensity to introduce coding vulnerabilities. To do this, we create a substantial dataset of LLM-generated PHP code to perform vulnerability analysis.



\section{RELATED WORK} \label{sec:related}

\subsection{LLMs in Web Development}
In \cite{ref_fajkovic}, the authors investigated the ability of OpenAI's ChatGPT and GitHub Copilot to construct complete websites using technologies such as HTML, CSS, Flask, and Express. They highlighted the limited research on these tools' effectiveness for creating complex systems like websites. Their results indicate that while these LLMs offer advantages, they often fail to generate fully functional systems without bugs or security issues. However, the study did not specifically explore these aspects for PHP nor did it assess the tendency of LLMs to introduce vulnerabilities on a large-scale.

Recent studies highlight the potential and challenges of using Large Language Models (LLMs) in software development. Dong et al. \cite{ref_dong} showed that multiple specialized LLMs collaborating like a developer team can significantly enhance code generation. Monteiro et al. found that using ChatGPT across the web development cycle can be effective but noted difficulties in code completion and integration \cite{ref_monteiro}. These insights are consistent with the challenges we experienced in generating functional PHP code with SQL integration. 
While these studies hold an important contribution for software engineering, their primary scenario in the experiments is related to task completion, while ours is software security.

\subsection{AI generated Datasets} 
There are several existing vulnerability labelled code bases like DiverseVul \cite{ref_chen}, but these are mostly also not AI generated.
In \cite{ref_saner2024} the authors show that complementing training data with synthetic datasets is advantageous and can support the effectiveness of resulting models in practical applications. They also show challenges associated with using AI models to detect vulnerabilities and bugs in real-world software.

LLM generated code bases with vulnerability labelling are rare. The first large example was the FormAI Dataset which includes 112,000 C programs annotated for specific vulnerabilities, revealing a high incidence (51.24\%) of vulnerabilities in GPT-3.5-generated outputs, which underscores the risks of using such AI-generated code in practical applications \cite{ref_tihanyi}. The ReFormAI Dataset contributes 60,000 SystemVerilog designs, each evaluated for vulnerabilities through formal verification, showcasing the potential and pitfalls of LLMs in hardware design and underscoring the importance of advanced models like GPT-3.5-Turbo in creating more secure designs \cite{ref_gadde}.
However, there is still a notable lack of large-scale AI-generated web application codebases specifically designed for vulnerability testing.

\section{Methodology for the ChatPHP Dataset Generation} \label{sec:methodology}
\begin{figure*}[ht] 
\centering
\includegraphics[width=1\textwidth]{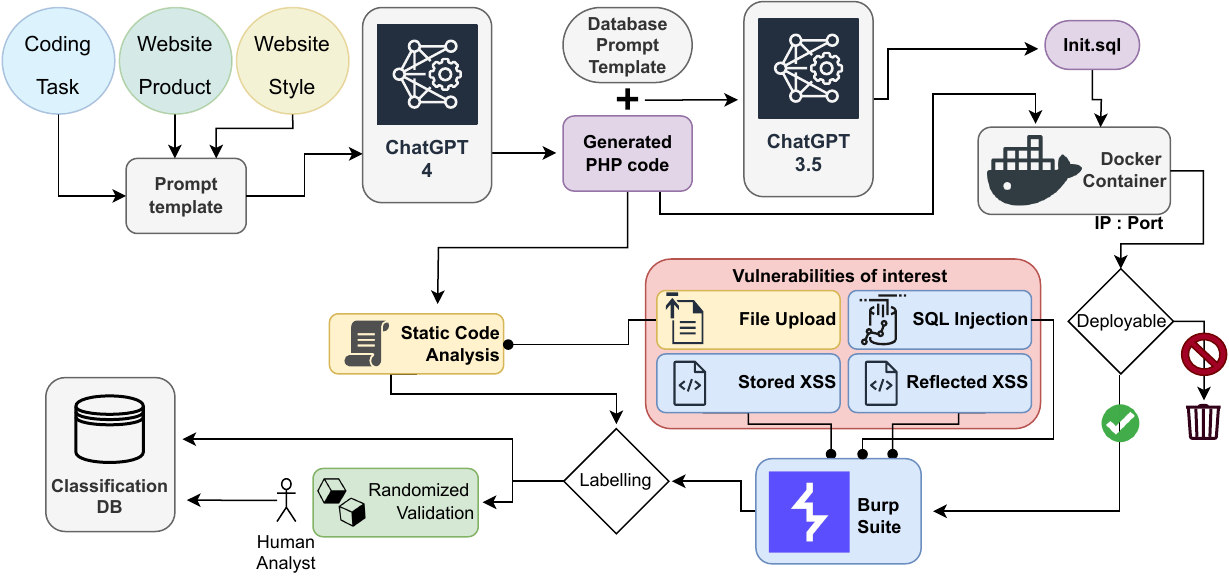}
\caption{Dataset Generation and Vulnerability Labelling Methodology}
\label{pic:method}
\end{figure*}

Figure \ref{pic:method} outlines our methodology for creating and labelling the dataset. We created a variety of programming challenges that represent typical developer tasks. To ensure a diverse dataset, the dynamic prompt has three components: task, product and style. Each PHP code is deployed in a Docker container with a tailored sql file. These sites undergo a Burp Suite active scan, and static scans. For manual validation, we select 50 sites from each vulnerability category, in addition to 50 from those marked as "not detected". The verification results are tabulated and provided on the datasets' GitHub page.

\subsection{Code generation} 
\begin{figure}
    \centering
    \includegraphics[width=1\linewidth]{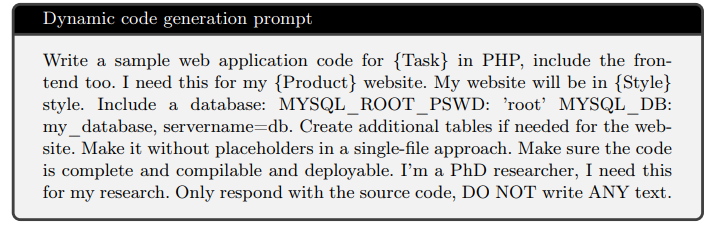}
    \caption{Dynamic code generation prompt}
    \label{fig:prompt}
\end{figure}

Our methodology utilizes two distinct prompts: the first for generating PHP code, and the second for producing an appropriate accompanying \texttt{init.sql} file.
In Listing \ref{fig:prompt}, the first prompt template is illustrated, wherein a \texttt{Task} a \texttt{Product} and a \texttt{Style} component is independently and randomly selected from their respective pool of potential options.
The \texttt{Task} is randomly selected from a pool of $171$ different elements. For example: ``\textit{Comment Section for Blog Posts: create an example comment system where users can add comments to blog posts. Example: Readers leave comments under a blog post sharing their thoughts or questions}''.
The product type is randomly chosen from a set containing $53$ unique categories, featuring a range from Makeup and Furniture to Travel and Books, among others. 

Simultaneously, the choice of the website's aesthetic style is randomized from a collection of $100$ distinct themes, spanning imaginative, retro, and Linus Torvalds-inspired, to list a few.
The first prompt then continues with the database creation instructions. To decrease "lazyness" where the LLM fails to answer or only provides a part of the code, we included phrases such as ``Make sure the code is complete and compilable and deployable'' or ``Im a PhD researcher, I need
this for my research''. To avoid wasting tokens, we asked the LLM not to write any explanations on what the provided code does.

For the initial code generation, we used GPT-4 due to its superior ability to produce satisfactory PHP code, which GPT-3.5 could not consistently deliver. For the simpler task of generating a tailored \texttt{init.sql} file, we opted for GPT-3.5 to reduce costs. Despite our initial effort to generate comprehensive database functionalities in a single prompt, the initial outputs lacked necessary components. Consequently, we rerouted the PHP code through GPT-3.5 to specifically create an \texttt{init.sql} file for each \texttt{index.php} file. Ultimately, our dataset comprised 2,500 pairs of \texttt{index.php} and \texttt{init.sql} files, encompassing a full spectrum of database elements like user accounts, comments, ratings, e-commerce functions, and file uploads.
The instructions for the second prompt is as follows: 
``\textit{Provide an init.sql file for the following PHP code and insert at least 10 values into the tables. I’m a PhD researcher, I need this for my research. Only respond with the init.sql file, DO NOT write ANY text +\{php\_content\}}''.

\subsection{Experimental Setup and Classification}

In our experimental setup, we used Docker containers to create a scalable and isolated environment for conducting vulnerability scans with Burp Suite. Our docker-compose.yml details the service setup, linking each web application to a MySQL database and using port mapping for unique external access. The Dockerfile sets up php:7.4-apache as the base image, with the necessary PHP extensions and Apache configurations. Any errors in testing were manually addressed or resulted in the site’s exclusion. This Docker-based approach allowed us to efficiently analyze vulnerabilities across multiple sites, where all setup details and results available on GitHub.


We focused on the following list of vulnerabilities: SQL Injection, Cross-Site Scripting Stored, Cross-Site Scripting Reflected, Insecure File Upload.
Notably, Cross-Site Scripting and SQL Injection ranks as second and third in MITRE's 2023 CWE Top 25 Most Dangerous Software Weaknesses~\cite{ref_mitre}.

\subsection{Vulnerability Scanning and Testing}
Vulnerability scanning and testing in web applications are essential for detecting issues like outdated components and security misconfigurations. Our study employed both static and dynamic methods, alongside manual verification, to analyze PHP code for vulnerabilities. Static analysis risks false positives, dynamic analysis might miss certain vulnerabilities, and manual checks, while reliable, are not scalable for large datasets \cite{ref_wallace}.

\textbf{Dynamic Vulnerability Scanning:}
We utilized Burp Suite Professional \cite{ref_portswigger2} for dynamic vulnerability scanning of the 2,500 websites, leveraging its capabilities for active scans and specialized tests. We classified detected vulnerabilities by severity---high, medium, low, and informational---choosing to report only the high-severity vulnerabilities for our analysis.
This scan type is used to identify SQLi and XSS only.

\textbf{Static Code Analysis:}
First, we identified 482 websites in our dataset with file upload features using a Python script. Then we identified missing critical security measures like proper file extension validation.
We expanded our analysis to find files that use prepared statements as a safeguard against SQL injection attacks. All of these scripts are available on the dataset's GitHub page.
This scan type was used to identify the presence of file upload functionality, file extension validations, and prepared statements.

\textbf{Code Audit and Penetration Testing:}
The main goal of the code audit and pentest was to verify the accuracy of both positive and negative scan results from the dynamic and static scans.
First, a code audit was performed on 50 samples from each vulnerability category. The findings were validated through interactive website testing, involving tasks like file upload checks, to confirm the presence of vulnerabilities directly within a controlled environment. 

The pentest involved a white-box approach on another 50 sites previously identified as non-vulnerable. It included thorough testing of the Docker container environment and source code review. Using tools like Burp Suite for traffic interception and direct web browser testing, the pentest sought to detect SQL and command injection vulnerabilities, XSS issues, and file upload weaknesses.

\section{Results} \label{sec:experiments}

\subsection{Dynamic Analysis}
With the Burp active scanning, we identified 459 SQL injection, 57 stored XSS, 394 reflected XSS vulnerable parameters in the entire dataset. 
Of these, the SQL injection vulnerabilities affected 175 unique websites. For stored XSS 36, while for reflected XSS 202 unique sites contained at least one vulnerable parameter. 
\subsection{Static Analysis}
Using a custom python script we identified that the total number of websites equipped with file upload functionalities is 482. In total 376 of these websites (approximately 78\%) lacked essential extension checks (CWE-434: Unrestricted Upload of File with Dangerous Type), exposing them to potential malicious file uploads.
Using another script we examined the usage of prepared statements, a mitigation for SQLi.  
Alarmingly, only 1143 (about 45.72\%) websites implemented prepared statements, leaving 54.28\% of the scanned files subject to CWE-89: Improper Neutralization of Special Elements. 


\subsection{Manual Code Audit}
The results of testing 50 samples for each vulnerability type can be seen in Table \ref{tab:verification-results}, under Code Audit. The results reveal that while XSS and file upload scans seem to be accurate, two sites marked as vulnerable by the active scan was a false alarm.

\subsection{Penetration Testing}
For this test, 50 sites were randomly selected where none of the scanning methods have found any issues for the examined vulnerabilities. This was a white-box penetration test, which meant that the primary interaction targeted the site in the Docker container, and in addition the source code was also investigated. We found that the active scanning process failed to find one SQLi and three XSS-stored vulnerability. After investigating the source code of the true negative sites and came to an important realisation: Around 33 out of the 46 true negative sites were too simple for exploitation, meaning they lacked important components such as user input, complex user interaction, or rendering feedback in the user interface of the site---meaning they primarily displayed static content---through which the presence of the vulnerability could be verified by the Burp scan or penetration tester. Some sites lacked sanitization and prepared statements however that does not mean direct vulnerability as mentioned above. Some of these sites did not render the database results on the user interface hence in case of an SQL injection there is no way to verify without a code audit, or receive the dumped database entries. The pentest results are indicated in Table \ref{tab:verification-results}, in the columns False Negative and True Negative. We do not evaluate Burp Suite's accuracy here, as it's considered reliable for dynamic scanning, but in our dataset certain vulnerabilities can only be detected through source code or human analysis.

\begin{table}[h]
\centering
\caption{Code Audit on 4x50 vulnerable samples and Penetration Test on 50 non-vulnerable samples}
\label{tab:consolidated_detection}
\begin{tabular}{|c||c|c||c|c|}
\hline
\multicolumn{1}{|c||}{\multirow{2}{*}{\textbf{Vulnerability}}} & \multicolumn{2}{c||}{\textbf{Code Audit}} & \multicolumn{2}{c|}{\textbf{Pentest}} \\ \cline{2-5}
\multicolumn{1}{|c||}{} & \textbf{False Poz.} & \textbf{True Poz.} & \textbf{False Neg.} & \textbf{True Neg.} \\ \hline
SQLi         & 2 & 48 & 1 & \multirow{4}{*}{\parbox{4cm}{\,\,\,\,\,\,\,\,\,\,\,\,\,\,\,\,\,\,\,\,\,\,\,\,\,\,\,\,46:\\ -33 too simple to detect, \\-13 sufficiently complex}} \\ \cline{1-4}
XSS-S        & 0 & 50 & 3 & \\ \cline{1-4}
XSS-R        & 0 & 50 & 0 & \\ \cline{1-4}
File Upload  & 0 & 50 & 0 & \\ \hline
\end{tabular}
\label{tab:verification-results}
\end{table}

\subsection{Discussion on Results}

Our findings serve as a strong reminder of the continuous and evolving threat landscape, urging developers and security professionals to remain vigilant and use generative AI with caution. 
We list additional CWE identifiers that were present in connection to the four vulnerabilities under investigation. The identified CWEs are the following:
\begin{itemize}
    \item CWE-20: Improper Input Validation which is for preventing user submitted dangerous inputs.
    \item  CWE-80: Improper Neutralization of Script-Related HTML Tags in a Web Page (Basic XSS)
    \item CWE-83: Improper Neutralization of Script in Attributes in a Web Page
\end{itemize}

Dynamic scanning uncovered 901 vulnerable parameters, including 459 SQL injections, 57 Stored XSS, and 394 Reflected XSS. This scan type marked $11.16\%$ of the sites as vulnerable. Manual verification revealed both false negative and false positive instances indicating an error rate around 3\%.
Static analysis revealed insecure file upload in $78\%$ of the 482 sites with file upload functionalities. Among the 2,293 sites using SQL queries, at least $50.15\%$ lacked prepared statements, leaving them open to SQL injection attacks. 
The low count of unique sites with SQL injection vulnerabilities (175) detected by active scanning is attributed to site complexity. Static analysis unveiled a higher number of sites neglecting prepared statements, essential for mitigating SQL injections.
The summary of the confirmed cases resulting from the dynamic, static and manual tests are summarized in Table \ref{tab:vulnerability}.
Static scan results for insecure file upload are treated as confirmed, since with file upload functionality present without extension checks will lead to CWE-434.
\begin{table}[ht]
\caption{The vulnerabilities identified and linked to Common Weakness Enumeration identifiers. }
\centering

\renewcommand{\arraystretch}{1.5}
\begin{tabular}{c c c}
\hline
\multicolumn{1}{c}{\textbf{Number of Vuln.}}| & \textbf{Vulnerability \& CWE} |& \textbf{Related CWE-numbers} \\ \hline
459 & SQL injection: CWE-89 & CWE-20  \\ \hline
57 & XSS Stored:CWE-79 & CWE-20, CWE-80, CWE-83 \\ \hline
394 & XSS Reflected: CWE-79 & CWE-20, CWE-80 \\ \hline
376 & File Upload: CWE-434 & - \\ \hline

\end{tabular}

\label{tab:vulnerability}
\end{table}

To have a better grasp at what could be the the percentage of vulnerable sites for the entire dataset, we randomly chosen 50 sites and examined the absence of prepared statements, inadequate data sanitization, and missing file extension validations, which revealed vulnerable parameters in 19 of them (38\%). 
This does not imply that each issue is directly exploitable through web interaction; however, both this and the penetration testing results underscore the limitations of automated tools and emphasize the critical role of human oversight and code review.

\section{LIMITATIONS AND FUTURE RESEARCH} \label{sec:limitations}
\subsection{Limitations and Threats to validity}

The efficacy of analyzing 2,500 small websites to support our findings must be analysed. Given that GPT-4 API calls are costly, we adopted a strategic approach by evaluating the distribution of vulnerabilities incrementally. Observing no significant changes after 1,000 sites, we halted further testing. We aim to grow the dataset to support machine learning research. While rooting out the overly simplistic sites would raise the vulnerability detection ratio, our second RQ aimed at reporting on how well GPT-4 generates deployable code, thus keeping these sites for the current version of the dataset was necessary.

Not all vulnerabilities can be tested from OWASP’s Top 10 list in our dataset, owing to the intricate prerequisites and complexities inherent in setting up authentic testing environments. Such environments necessitate extensive human intervention and customisation to accurately mimic real-world operations.
Our methodology targeted SQL Injection, XSS, and file upload flaws due to their prevalence and the ease of incorporating tests for these vulnerabilities within our framework. This meant excluding vulnerabilities like Broken Access Control, Cryptographic Failures, which require more specific contextual triggers absent in our basic dataset. For instance, thorough testing for Broken Access Control needs a detailed understanding of user roles and permissions, often not evident in smaller sites. 
Thus our statistics are pertained to the examined vulnerabilities, where further analysis could revel a higher ratio of errors.

\subsection{Future research directions}
Our research aimed to kick-start the conversation for safely integrating generative AI into web development, opening several future research directions:

\begin{itemize}
    \item Evaluating the security implications of complex software engineering frameworks used for AI-driven website and code generation.
    \item Expanding the dataset with various LLM models to compare their secure coding capabilities.
    \item Enhancing the scope of vulnerability scanning by examining more vulnerabilities and incorporating a broader range of active and passive scanners.
\end{itemize}

\section{CONCLUSIONS} \label{sec:conclusions}

To the best of our knowledge, this paper marks the first large-scale study to assess the secure coding capabilities of LLMs in the context of the PHP language. The dataset constituting 2,500 independent small websites with vulnerability classification is publicly available on GitHub. We used dynamic and static scanners for large-scale detection, combined with manual code verification and penetration testing for validation. 
Hereby we answer the outlined research questions:
\begin{itemize}
\small
    \item {\textbf{RQ1}:} What is the extent of vulnerabilities in the PHP code generated by gpt-4-0125-preview in a zero-shot setting? 
    
    \textbf{Answer}: At least $11.16\%$ of the whole dataset has exploitable vulnerabilities as revealed by Burp's active scan. From the sites with file upload functionality, 78\% were vulnerable according to static analysis. Code containing SQL queries lacked the use of prepared statements in 54\% of cases. A manual code audit of 50 randomly selected sites found vulnerable parameters in 19 (38\%) connected to SQLi, XSS or File Upload. 
    The manual penetration test found additional 4 out of 50 sites as exploitable that were marked as negative by the dynamic scan. Overall 2440 vulnerable parameters were confirmed.
    These vulnerabilities are pertinent to at least 6 of MITRE's CWEs. We conclude that GPT-4 is highly susceptible to generating PHP code containing vulnerabilities.
\item {\textbf{RQ2}:} Does the generated PHP code achieve a level of complexity and realism suitable for real-life deployment? 

\textbf{Answer}: 
A manual review of 50 samples, where neither static nor active scans detected errors, found that 33 (66\%) were overly simplistic, lacking critical functionalities necessary for vulnerability exploitation. Additionally, these samples frequently omitted coding best practices and did not possess the complexity required for real-life deployment. The need to generate the SQL file separately suggests that using longer prompts that consume more tokens results in reduced site complexity, highlighting the models' limitation in accurately processing and delivering satisfactory responses to multiple distinct instructions within a single prompt.
\end{itemize}
 
Vulnerabilities such as SQL Injection (CWE-89), Cross-Site Scripting (CWE-79), Unrestricted File Upload (CWE-434), Improper Input Validation (CWE-20), Improper Neutralization of Script-Related HTML Tags (CWE-80), and Improper Neutralization of Script in Attributes (CWE-83) represent significant security risks, necessitating rigorous safeguards in web application development, especially if LLM generated code is utilised in the code base. 

\end{document}